\documentclass{ieeeaccess}

\usepackage[utf8]{inputenc}
\usepackage{cite}
\usepackage{amsmath,amssymb,amsfonts}
\usepackage{algorithmic}
\usepackage{graphicx}
\usepackage{textcomp}
\usepackage{comment}
\usepackage{listings}
\usepackage{siunitx}
\usepackage{color}
\usepackage{enumitem}
\usepackage{booktabs}
\usepackage{gensymb}

\renewcommand{\thesubsection}{\thesection.\Alph{subsection}}

\newcommand{\TitleText}{High-Resolution Waveform Capture Device on a Cyclone-V FPGA}

\newcommand{\TDL}{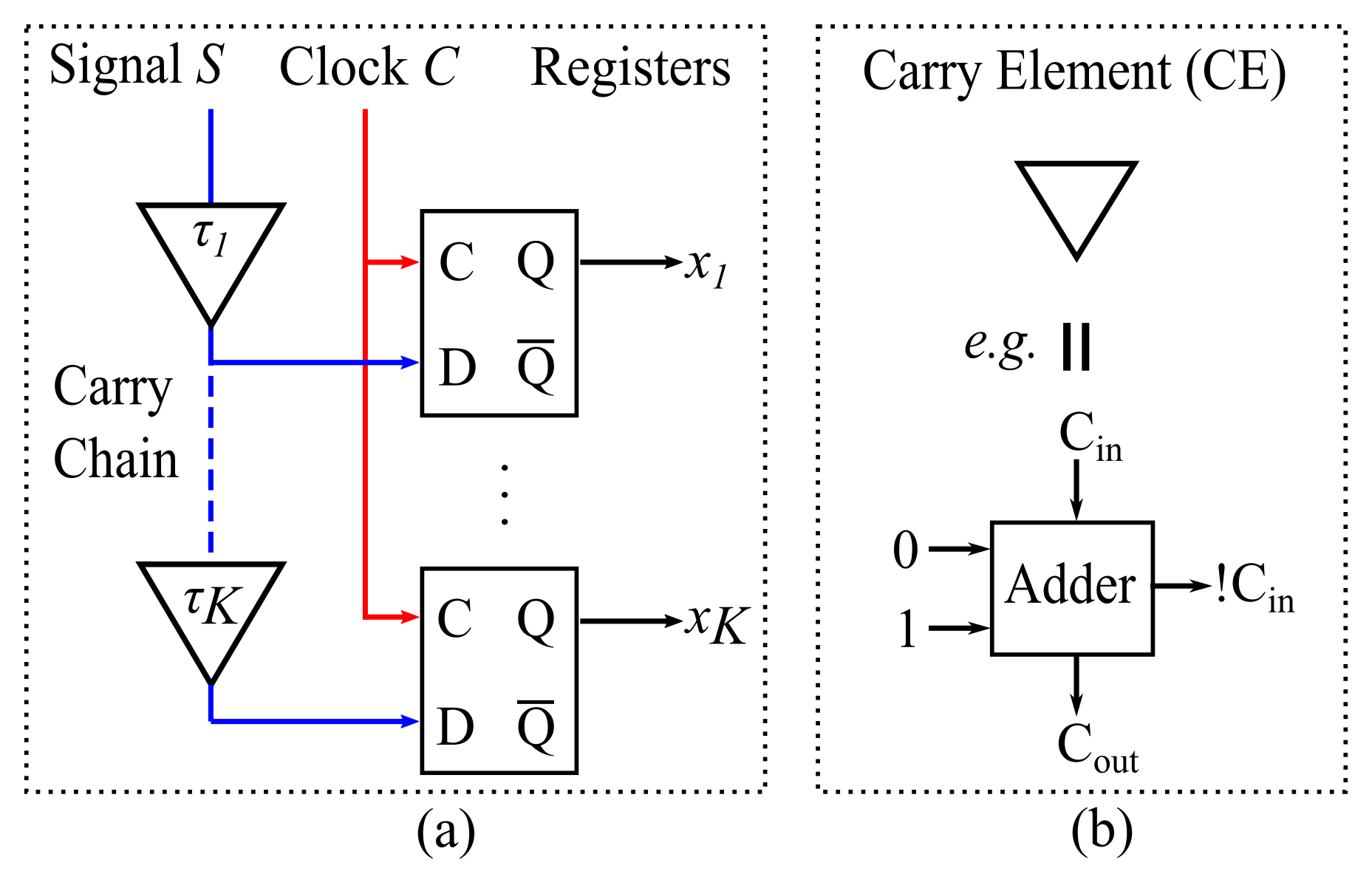}
\newcommand{\Disc}{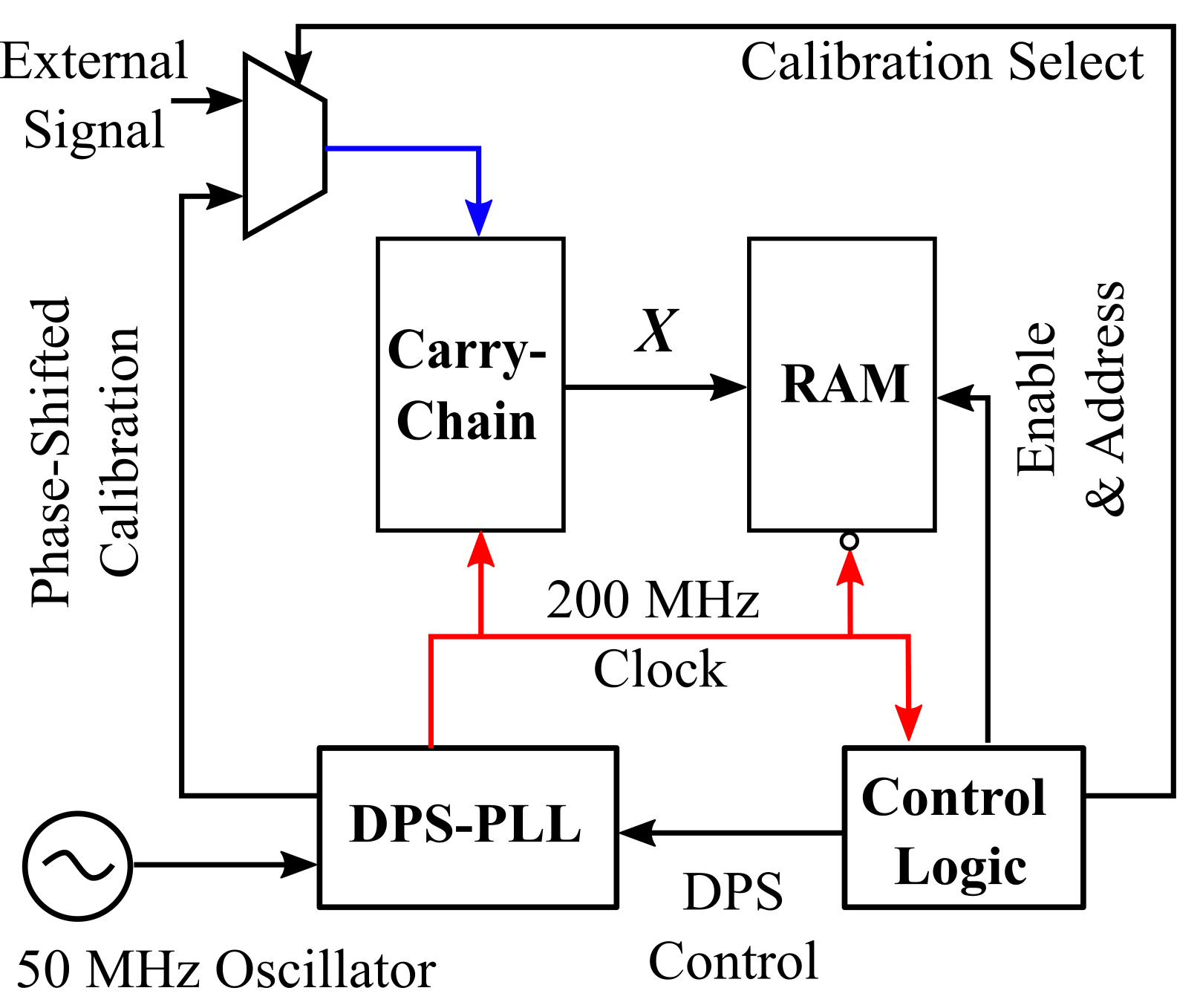}
\newcommand{\FPGA}{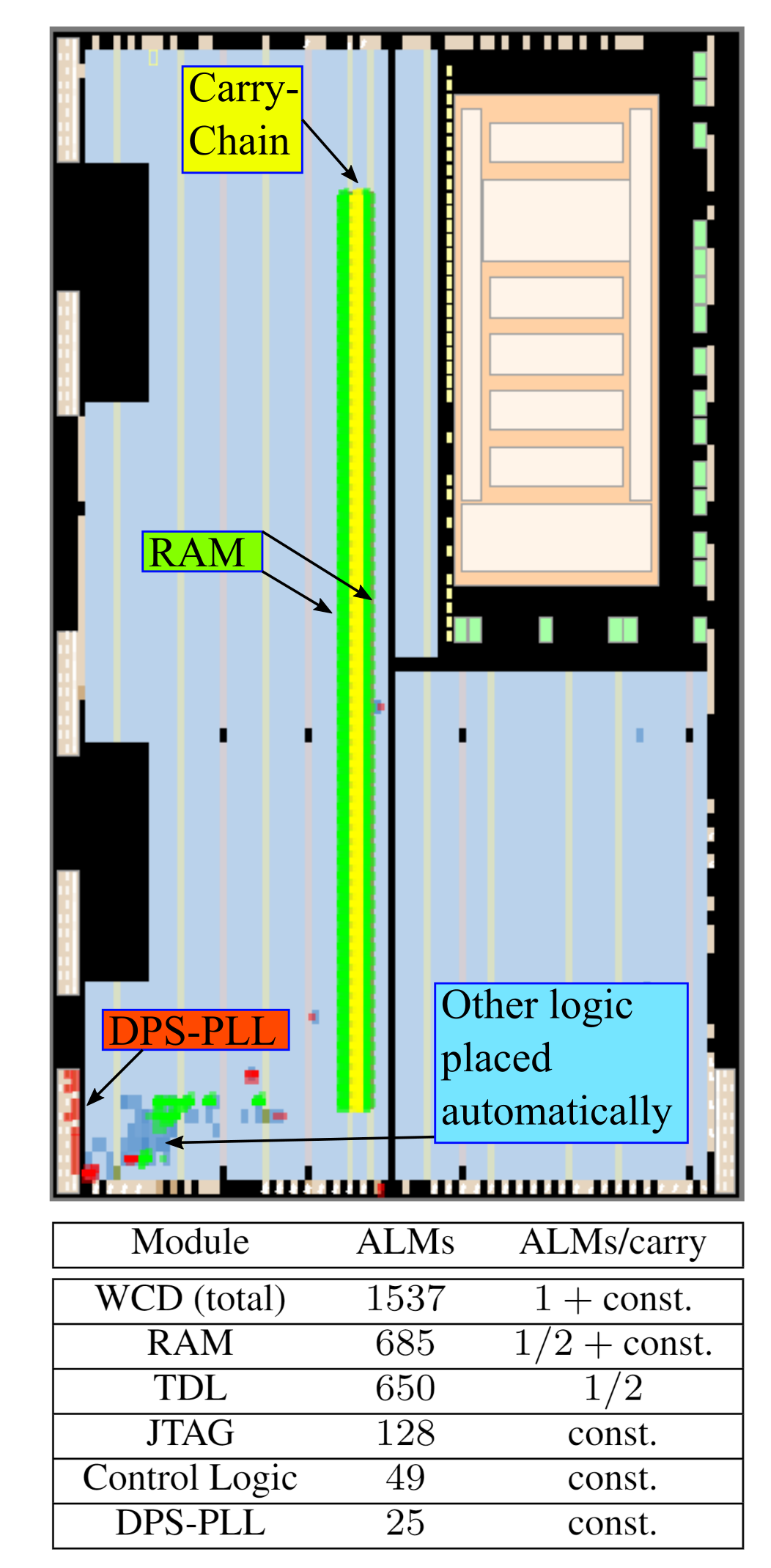}
\newcommand{\Data}{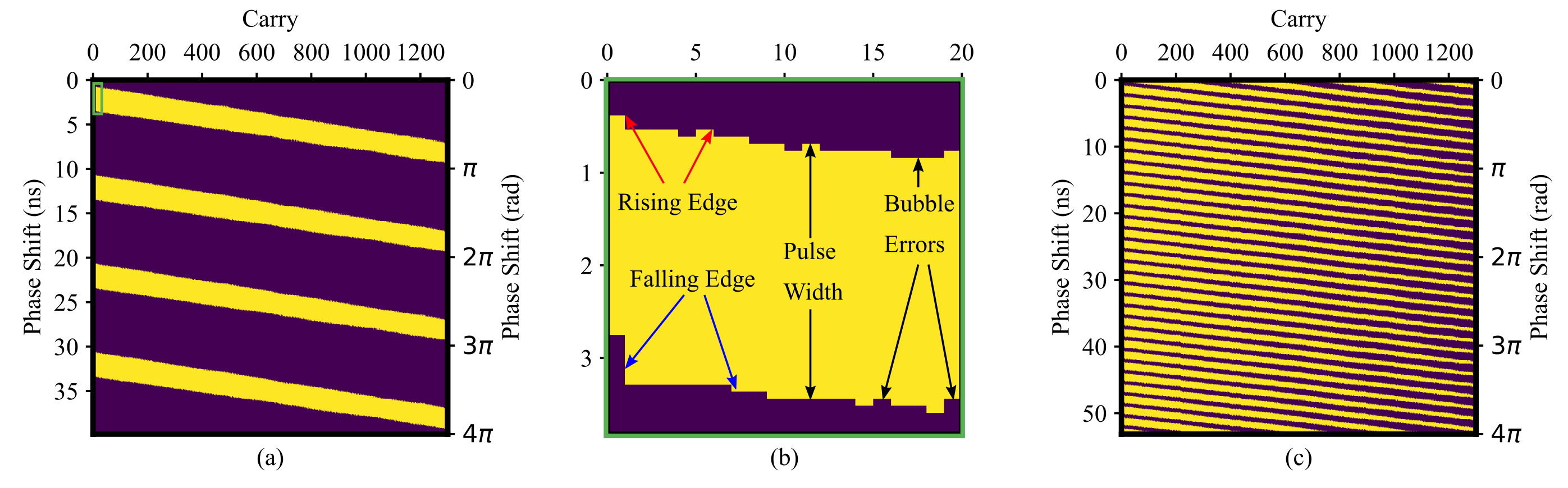}
\newcommand{\Calib}{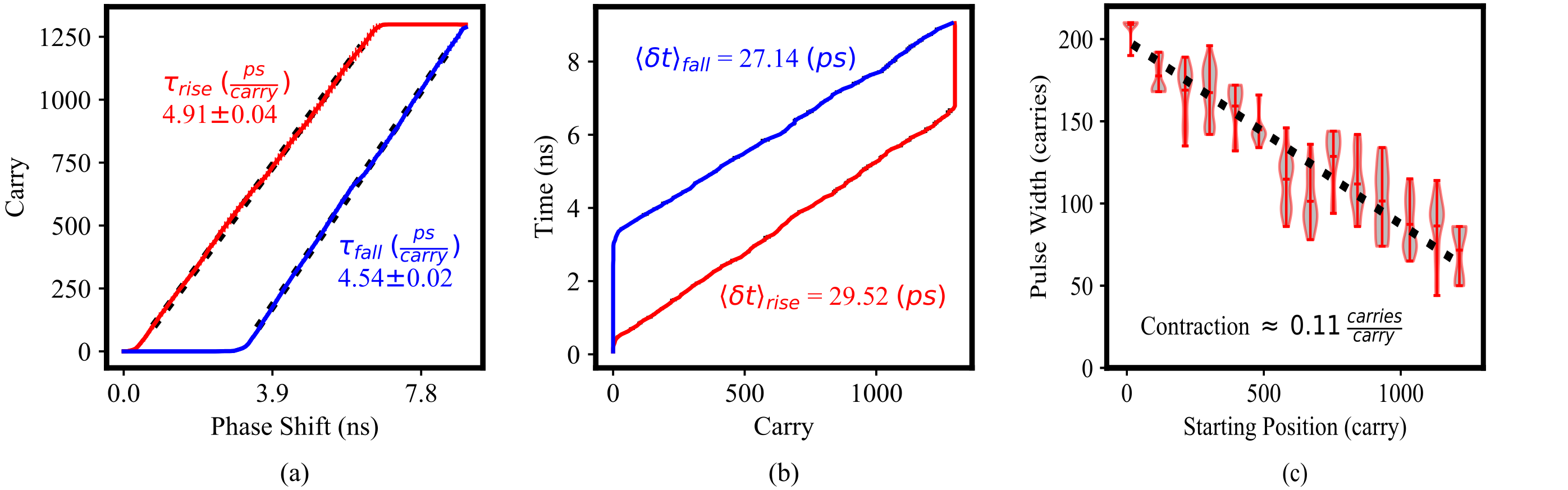}
\newcommand{\EC}{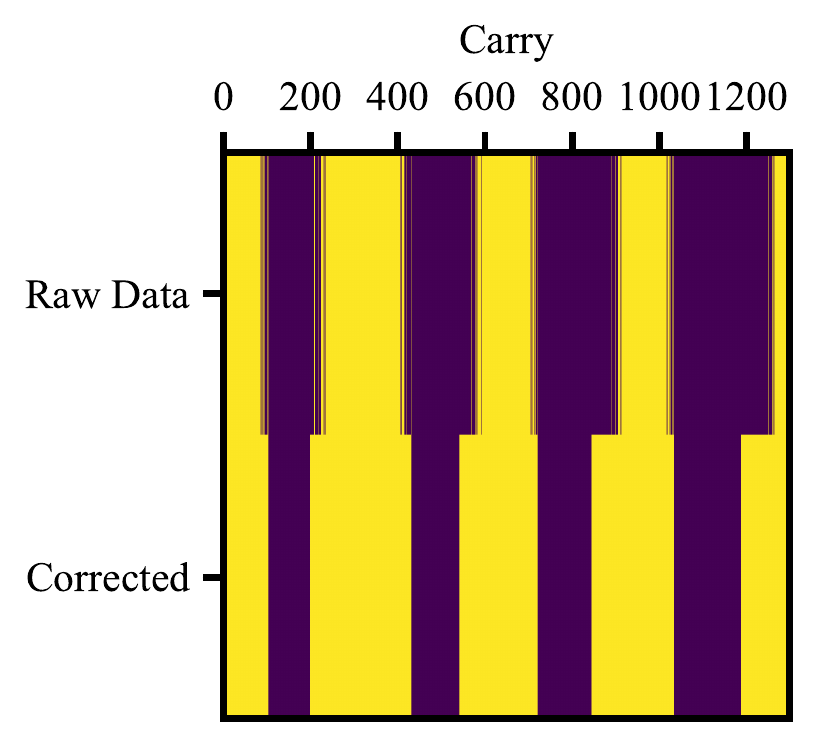}
\newcommand{\RO}{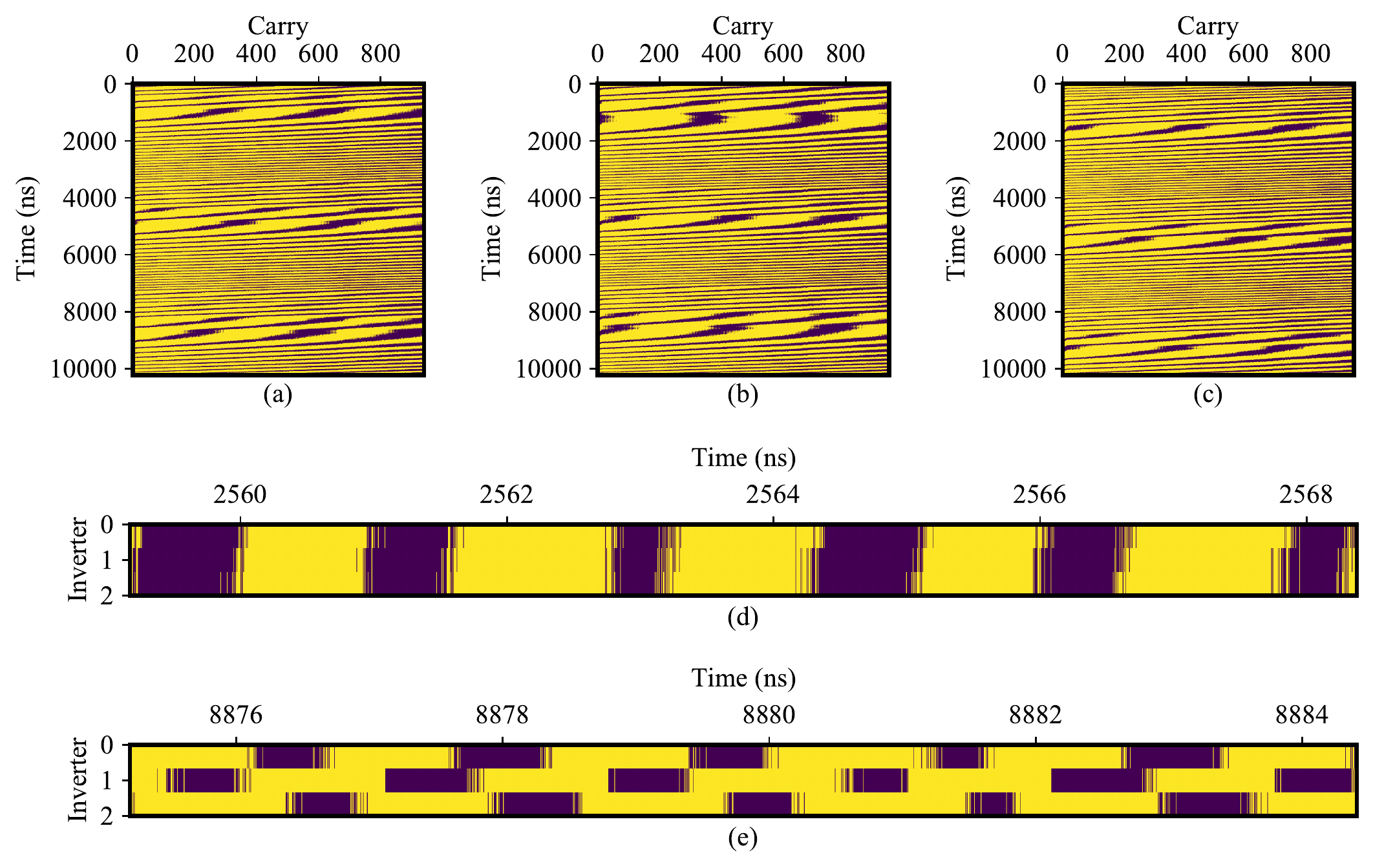}
\newcommand{\LE}{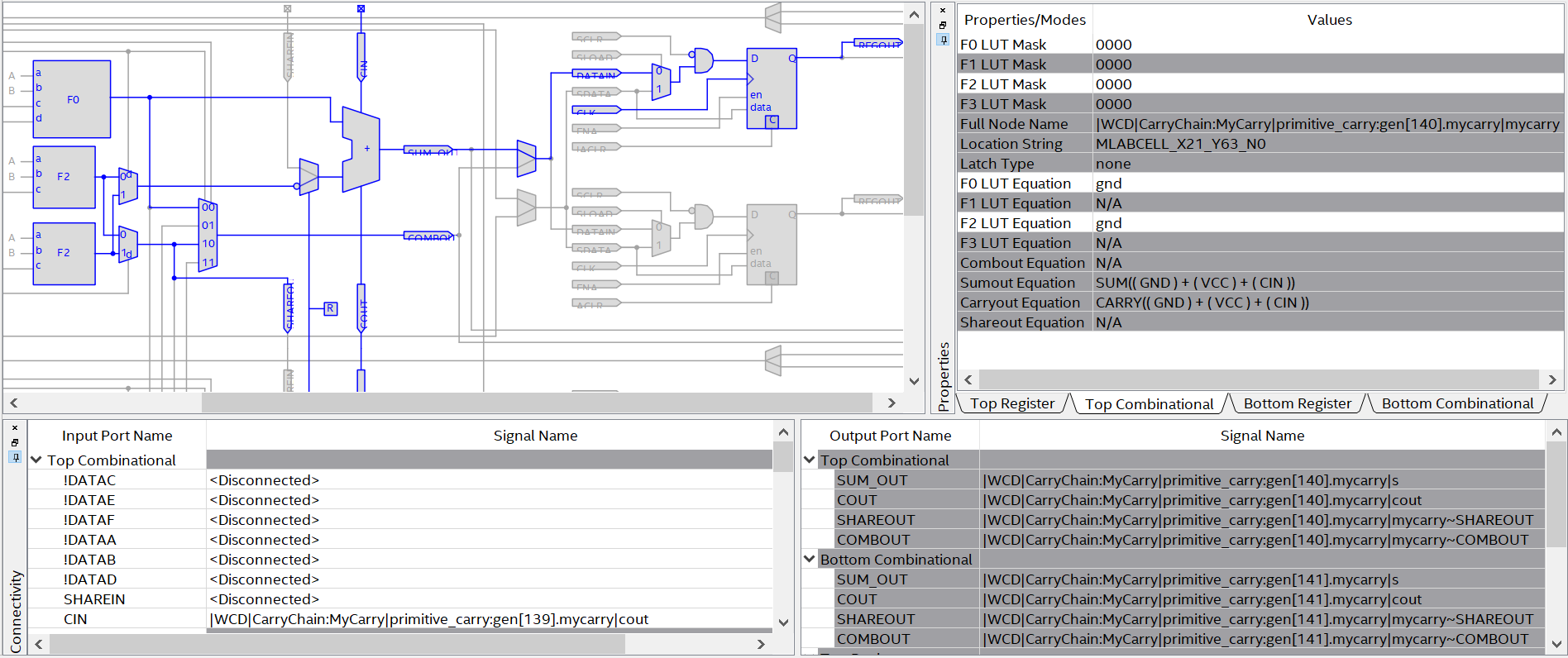}

\newcommand{\SingleShotRise}{$29.52$ ps}
\newcommand{\SingleShotFall}{$27.14$ ps}

\newcommand{\LABCarryRiseTime}{$6.35\pm0.15$ ps}
\newcommand{\LABCarryFallTime}{$5.60\pm0.12$ ps}
\newcommand{\CarryRiseTime}{$4.91\pm0.04$ ps}
\newcommand{\CarryFallTime}{$4.54\pm0.02$ ps}
\newcommand{\LogicHighTime}{$0.240\pm0.002$ ns}

\definecolor{codegreen}{rgb}{0,0.6,0}
\definecolor{codegray}{rgb}{0.5,0.5,0.5}
\definecolor{codepurple}{rgb}{0.58,0,0.82}
\definecolor{backcolour}{rgb}{0.95,0.95,0.92}

\lstdefinestyle{mystyle}{
    backgroundcolor=\color{backcolour},   
    commentstyle=\color{codegreen},
    keywordstyle=\color{magenta},
    numberstyle=\tiny\color{codegray},
    stringstyle=\color{codepurple},
    basicstyle=\footnotesize,
    breakatwhitespace=false,         
    breaklines=true,                 
    captionpos=b,                    
    keepspaces=true,                 
    numbers=left,                    
    numbersep=5pt,                  
    showspaces=false,                
    showstringspaces=false,
    showtabs=false,                  
    tabsize=2
}

\usepackage{hyperref}
\def\BibTeX{{\rm B\kern-.05em{\sc i\kern-.025em b}\kern-.08em
    T\kern-.1667em\lower.7ex\hbox{E}\kern-.125emX}}
\begin{document}
\history{UNPUBLISHED April 2021}
\doi{N/A}

\title{\TitleText{}}
\author{\uppercase{Noeloikeau Charlot}\authorrefmark{1},
\uppercase{Daniel J. Gauthier\authorrefmark{2}, and Andrew Pomerance\authorrefmark{3}}}
\address[1,2]{Ohio State University, Department of Physics, 191 West Woodruff Ave, Columbus, OH 43202, USA, (e-mail: charlot.5@osu.edu, gauthier.51@osu.edu)}
\address[3]{Potomac Research, LLC, 2597 Nicky Lane Alexandria, VA 22311, USA (e-mail: andrew.pomerance@gmail.com)}

\tfootnote{This work was supported by the Department of the Army through award number W31P4Q-20-C-0003. A preliminary patent was filed based on this work under Application No. 63/177,435 for 5-PS-RESOLUTION WAVEFORM-CAPTURE-DEVICE ON A CYCLONE V FIELD-PROGRAMMABLE GATE-ARRAY WITH DYNAMIC PHASE-SHIFTING.}

\markboth
{Noeloikeau Charlot \headeretal: \TitleText{}}
{Noeloikeau Charlot \headeretal: \TitleText{}}

\corresp{Corresponding author: Noeloikeau Charlot (e-mail: charlot.5@osu.edu).}

\begin{abstract}
We introduce the waveform capture device (WCD), a flexible measurement system capable of recording complex digital signals on trillionth-of-a-second (ps) time scales. The WCD is implemented via modular code on an off-the-shelf field-programmable gate-array (FPGA, Intel/Altera Cyclone V), and incorporates both time-to-digital converter (TDC) and digital storage oscilloscope (DSO) functionality. The device captures a waveform by taking snapshots of a signal as it propagates down an ultra-fast transmission line known as a carry chain (CC). It is calibrated via a novel dynamic phase-shifting (DPS) method that requires substantially less data and resources than the state-of-the-art. Using DPS, we find the measurement resolution - or mean propagation delay from one CC element to the next - to be \CarryRiseTime{} (\CarryFallTime{}) for a pulse of logic high (low). Similarly, we find the single-shot precision - or mean error on the timing of the waveform - to be \SingleShotRise{} (\SingleShotFall{}) for pulses of logic high (low). We verify these findings by reproducing commercial oscilloscope measurements of asynchronous ring-oscillators on FPGAs, finding the mean pulse width to be \LogicHighTime{} per inverter gate. Finally, we present a careful analysis of design constraints, introduce a novel error correction algorithm, and sketch a simple extension to the analog domain. We also provide the Verilog code instantiating the our design on an FPGA in an Appendix, and make our methods available as an open-source Python library at \href{https://github.com/Noeloikeau/fpyga}{https://github.com/Noeloikeau/fpyga}.
\end{abstract}


\titlepgskip=-15pt

\maketitle

\section{Introduction}
\label{IntroductionSection}
\PARstart{M}{easuring} time-dependent digital waveforms with accuracy and precision is crucial in many fields of science and engineering, with applications ranging from wireless communication systems and GPS to medical imaging and PET scans\cite{TimingGenerator,ImagingTutorial}. Depending on the use-case, two technologies are routinely employed: time-to-digital converters (TDCs)\cite{Review2019}, and digital storage oscilloscopes (DSOs)\cite{FPGA-DSO}. DSOs measure the history of a signal as it varies over time (its waveform), with analog-input (AI-DSO) and digital-input (DI-DSO) commonly available. TDCs measure the time interval between digital events, akin to a stopwatch. Historically, the two technologies have been composed of different application-specific integrated-circuits (ASICs).

Recently however, field-programmable gate arrays (FPGAs) have become an attractive alternative to ASICs for many applications. FPGAs can act as a `one-size fits all' solution due to their ability to implement nearly any circuit by reconfiguring their hardware using only code and software.

Software-defined DI-DSOs based on standard FPGA logic elements are usually found in modern FPGA electronic design suites for debugging user-defined circuits. The sample rate of these tools is usually limited to the range of 200-300 MSamples/s because of the difficulties of registering the digital signal at a specific location in the circuit and storing the value in memory.

On the other hand, FPGA-based TDCs have flourished in recent years using existing FPGA resources in a non-standard configuration. One approach is to use the carry-chain (CC) elements built into each logic element to realize a high-speed transmission line for a digital signal. Here, it is assumed that the pulse width is longer than the transmission line delay time and the goal is to register the edge of the pulse. The CC elements have been especially designed for high-speed operation of on-chip adders, and a digital signal can be passed from one CC element to the next on the picosecond time scale. Each CC element has an associated register, which can be triggered by a global clock signal to freeze the digital waveform contained in the transmission line. The digital output of the TDC is a word that encodes the time difference between the edge and the reference clock.

Using this approach, TDCs have been realized with few ps resolution \cite{CycloneV2ps,Xilinx4ps,Xilinx5ps}. However, calibrating the FPGA-TDCs is challenging because it is difficult to generate ps-scale signals on the chip. Also, signal distortions have to be corrected that arise from small timing variations of the signal from the carry-chain to the register, and from digital pulse contraction or expansion due to differences in the propagation speed of rising or falling edges along the transmission line \cite{KeyIssues}.

FPGA-TDCs have accounted for their shortcomings with a variety of calibration procedures, the goal of which is to obtain accurate and precise measurements from raw data (\textit{i.e.}, data that has not been post-processed). In the case of a TDC, the goal is to obtain a linear relation between the timing between an edge of an input digital signal and a reference clock edge. Many calibration methods have been proposed, such as statistical code density testing \cite{Review2019}, wave-union \cite{WaveUnion}, direct-to-histogram \cite{DirectToHist}, multi-phase sampling \cite{Multiphase}, variable clock generation \cite{NambaClock1,NambaClock2,NambaClock3}, and dynamic reconfiguration \cite{DynamicDelayLoop,DynamicReconfigTDC}.

Many of these calibration techniques require complex circuit architectures, high logic utilization, and large data sets. This is due to the aforementioned difficulty in generating ps-scale time intervals on FPGAs, which translates into a lack of control over the calibration signal, its time of application, or the measurement time. Clever solutions to this problem have been devised, such as the statistical code density test\cite{Review2019}, which uses signals that are randomly distributed with respect to the reference clock. This method generates a uniform sampling of time intervals, but requires a large amount of data to obtain statistically relevant results. For these reasons, the calibration of an FPGA-TDC is often the most challenging and tedious part of the design. The result of the calibration process is a compressed, corrected, or otherwise encoded signal using auxiliary circuits such as edge detectors, thermometer-to-binary encoders, and bubble-error correctors \cite{Review2019,XilinxMissingCodeFree,BubbleCorrection}. 

The design of our waveform capture device (WCD) is similar to these FPGA-TDC designs, but it uses long CCs to capture a digital waveform without encoding it. As a result, we are able to directly measure the dynamics of arbitrary digital signals - not just timestamp pulse edges - with little overhead. We demonstrate both TDC and DI-DSO capabilities with precise measurements of complex GHz frequency waveforms entirely within an FPGA.

We calibrate our device using a novel method we refer to as dynamic phase-calibration (DPC). The method is applicable to FPGA-TDCs starting with the Cyclone V and Xilinx 7 families of FPGAs, with continued support for the most recent chip families. DPC relies on a technology known as dynamic phase-shifting (DPS) \cite{AlteraPLL, XilinxPLL}, which functions by adjusting the phase of a stable on-board clock known as a phase-lock loop (PLL). DPS reliably generates precise ps-scale time-intervals without re-configuring the FPGA. As a result, many of the former difficulties and complications associated with TDC calibration are eliminated.

As with any temporal measurement system, we are limited by the precision and accuracy of the reference clock used to trigger the waveform capture registers. Here, we use the DE10-Nano-SoC demonstration board manufactured by Terasic, which has a Siliconix phase locked loop model \# Si5350C driven by a 27-MHz crystal oscillator. The Siliconx device has a typical period and cycle-to-cycle jitter of 40 ps and 50 ps, respectively, over 10,000 oscillations \cite{OscillatorDeviceDatasheet}. As we show, this jitter is the primary source of error in our measurement system, however the oscillator can easily be replaced by a higher-stability device as needed.

Our unique contributions are as follows:

\begin{itemize}
    \item We introduce a novel FPGA-based measurement system using CCs that is capable of directly capturing digital waveforms on ps time-intervals, incorporating the core functionality of DI-DSOs and TDCs while using a simpler architecture.
    \item We introduce a novel calibration method using dynamic phase-shifting that is applicable to current DI-DSOs and TDCs on FPGAs, replacing the large logic overhead and data requirements of the state-of-the-art with a straightforward velocity calculation.
    \item We validate these claims through several case studies on PLLs and ring-oscillators, finding agreement with previous studies. We go on to report measurements of FPGA logic with unparalleled precision, study complex digital waveforms, and identify novel error correction techniques to reduce signal distortions in CCs on the Cyclone V.
\end{itemize}

The paper is organized as follows. In Sec. \ref{WCD}, we describe the design and implementation of the WCD. In Sec. \ref{Calibration}, we examine the raw data and perform calibrations and error correction. In Sec. \ref{Results}, we validate the design using Ring Oscillators as a case-study. We conclude in \ref{Conclusion} with a discussion and future work.

All experiments are performed on DE10-Nano SOCs hosting Cyclone V 5CSEBA6U23I7 FPGAs. Data is collected over the JTAG interface using the open-source `fpyga' Python package maintained by the primary author at \href{https://github.com/Noeloikeau/fpyga}{https://github.com/Noeloikeau/fpyga}, which also interfaces with the Quartus CAD software to implement all location and timing constraints. Our design uses low-level control over the FPGA primitives. These are poorly documented\cite{CycloneIICarryPrimitive}; as such, we provide the Verilog code instantiating our design in the Appendix.

\section{WCD Design and Operation}
\label{WCD}
In this section, we begin by summarizing the working principles and overall design of the WCD before presenting a careful analysis of the design constraints and implementation in the following subsection. In short, the WCD is a digital buffer that continuously measures the history of an input signal, and DPC is a velocity calculation that measures the distance travelled down this buffer after a phase-shift.

The core component of the WCD is a tapped delay line (TDL), described in Fig. \ref{FTDL}. The TDL is composed of a CC used to buffer an input signal and registers used to capture its waveform on the edge of a clock. The CC is composed of a chain of $K$ carry-elements (CEs) that pass the signal to the next CE after some time-delay $\tau_{k}$. Each CE is composed of a redundant logic element such as an adder with fixed inputs whose sole purpose is to route the input signal down the special high-speed transmission lines that FPGAs reserve for carry operations. As in Fig. \ref{FTDL} (b), the signal also passes from each carry to a corresponding register. Each register stores the signal at that point in the chain as a bit $x_{k}$ when a global clock $C$ goes high, effectively capturing the waveform.

\Figure[!htb](topskip=0pt, botskip=0pt, midskip=0pt)[width=3.3in]{\TDL}
{Primary building-blocks of the WCD. (a): Tapped Delay Line (TDL) composed of multiple registers and (b): Carry Elements (CEs). Note: the CE is not truly an adder, see the Appendix.\label{FTDL}}

The concatenation of the register outputs forms a $K$-bit digital word $\textbf{X}=\{x_{1},...,x_{K}\}\in\{0,1\}^{K}$ directly encoding the captured waveform. Suppose each of the $K$ carries acts as a nominal time-delay $\tau$. It takes $K\tau$ time for $S$ to travel down the CC. Further suppose that the clock $C$ triggers all registers in the TDL simultaneously at time $t$. The TDL captures $S$ backward in time from $t$ in $K$ increments of time $\tau$. Hence, the principle of waveform capture is
\begin{equation}
    \label{WC}
    \textbf{X}(t)=\theta(\{S(t - \tau),...,S(t-K\tau)\}),
\end{equation}
where $\theta: [S_{min},S_{max}]^{K}\rightarrow\{0,1\}^{K}$ is an element-wise Booleanization operation representing the digitization of the signal performed by the registers.

\Figure[!htb](topskip=0pt, botskip=0pt, midskip=0pt)[width=3.0in]{\Disc}
{WCD block diagram.\label{FDisc}}

Figure \ref{FDisc} shows the WCD and DPS-calibration block diagram. The DPS-PLL (dynamic phase-shifting phase-lock loop), RAM, and TDL are governed by control logic, and the entire WCD is clocked continuously by a 200 MHz PLL output $C$. When not calibrating, the WCD captures the last $K\tau=(1300)(\approx 5\text{ ps})=6.5\text{ ns}$ of the waveform of $S$ every $T=5$ ns on the positive edge of $C$. This provides redundancy and ensures no portion of the waveform is lost. The data is then pushed to RAM as a $K=1300$ bit digital word on the negative edge of $C$, chosen to avoid read/write conflicts with the TDL. This repeats until $512$ periods or $2.56$ $\mu s$ have passed (unless otherwise stated), and the data is transferred onto a computer. We stress that there is no dead-time on the WCD when not calibrating, and the device acts as a DI-DSO / TDC hybrid.

When calibrating, the signal $S$ is switched by the control logic to a second PLL output whose frequency is a ratio of 200 MHz. $S$ is then shifted in phase relative to $C$ in increments $\Delta t=78$ ps by engaging the phase-shifting ports. When the DPS operation finishes after a minimum of two 50 MHz cycles of the external oscillator, the memory address is updated, the write enable signal is sent by the control logic to the RAM, and the TDL captures $S$ on the rising edge of $C$. The net phase shift on the signal at this time is $n \Delta t$, where $n$ is the memory address at which $X$ is stored on the negative edge of $C$. The process repeats until 512 increments have occurred, corresponding to a phase-shift of $4\pi$, and the difference in carry positions of the edge of the signal at each phase-shift are used to calibrate the device.

\subsection{Constraints and Implementation}

The WCD is fundamentally limited by the mean carry time $\tau$. This is the smallest time-interval between which the signal is registered, and is often referred to as the resolution \cite{MetricsThesis}. As we show, this value is approximately $\tau \sim 5$ ps for the Cyclone V architecture. The resulting design space is then constrained by the length of the carry chain $K$ and the period of the capture clock $T$. 

For the WCD to continuously measure $S$, the time required for the signal to travel down the chain $K\tau$ must be greater than or equal to the time between captures $T$. Otherwise, portions of the waveform are lost in between clock intervals. Hence, we require
\begin{equation}
    \label{constraint}
    K\tau \geq T.
\end{equation}
Further constraints on $K$ arise due to the device-dependent properties of the FPGA architecture. For the Cyclone V DE10 Nano, the maximum CC length is $K=1740$, which spans the height of the chip. Because the carry must enter from the top of the chain, this sets the maximum possible buffer time $K\tau$. As we show below, there also exist variations in the rise and fall times of the carry elements $\tau_{rise}>\tau_{fall}$. Consequently, a pulse traveling down the chain experiences contraction due to the falling edge (tail) catching up to the rising edge (head) of the pulse. This further limits $K$ according to the tolerable level of pulse-shortening. That is, if the minimum time between pulses to be resolved is $T_{min}$, then $K(\tau_{rise}-\tau_{fall}) < T_{min}$.  In practice we find $K\leq 1500$ prevents total degradation of sub-nanosecond pulses. 

There are also limitations on the signal $S$ entering the TDL via the internal FPGA routing. If $S$ is generated external to the FPGA, it must first pass from a pin to an I/O buffer. Further, $S$ must enter the TDL through an initial copy-gate before being carried to the CC, as it cannot be fed directly into a carry port. Most importantly, the path the signal takes through the CC must be set as a false path in the timing analyzer. This is because the travel time through the CC imposes false timing constraints that cannot be met by the compiler and must be ignored to avoid critical timing failures. 

Timing constraints are primarily the skew - or difference in travel time of the clock signal - to each register, as well as the travel time required for the data to be transferred from the TDL to RAM. These are minimized by routing the on-chip PLL governing waveform capture through a global clock fabric minimizing the skew, as well as placing the RAM and TDL directly adjacent to each-other as shown in Fig. \ref{FFPGA}. We find that these techniques allow for continuously registering the CC without dead-time up to a $200$ MHz PLL output, yielding a time between captures of $T=5$ ns. This strikes a balance between the nominal minimum period $T=1.43$ ns derived from the maximum frequency of 700 MHz for Cyclone V PLLs, and the approximate maximum buffer time of $7.5$ ns due to a $K=1500$ CC.

DPS calibration is limited by the minimum possible dynamic phase shift $\Delta t=78$ ps for a Cyclone V, given by $1/8$ the maximum frequency of $1600$ MHz for the voltage controlled oscillator (VCO) governing the PLL. This phase-shift must be minimized to maximally populate the number of edge positions recorded at each carry during calibration. Dead-time arises due to the process of engaging the auxiliary ports on the PLL enabling DPS. The first two ports need only be set once, and specify the desired PLL output to phase-shift and the direction to phase-shift; in this study these are set to static values so as to shift the calibration signal in the positive direction. However, to begin DPS, the last port must be asserted logic high for two clock cycles of the external 50 MHz free-running external crystal oscillator onto which the PLL locks. This process unlocks the PLL to the reference oscillator and the re-locking procedure is asynchronous and device-dependent; in practice, we find it ranges from the minimum of two clock cycles to several dozen depending on the FPGA and operating conditions. For further details on engaging DPS in the Cyclone V and Xilinx 7, consult Ref.~\cite{AlteraPLL} and \cite{XilinxPLL}, respectively. 

\Figure[!htb](topskip=0pt, botskip=0pt, midskip=0pt)[width=2.5in]{\FPGA}
{Resource allocation for the $K=1300$ carry-element WCD on a Cyclone V used in this study. Highlighted are the primary modules including the Carry-Chain/TDL, RAM, and the DPS/PLL.\label{FFPGA}}

Finally, the uncertainty on the timing of the waveform is limited by the relative timing errors on the two outputs of the PLL, and by the jitter of the Siliconix crystal oscillator that drives them. The Cyclone V reference manual lists the maximum jitter on the PLL over 15 standard deviations as $300$ ps and on the phase shift as $50$ ps \cite{AlteraPLL}, though these values are highly conservative. As we show in the next section, the actual timing error - quantified by the single-shot precision, or the root-mean-square timing error over the full CC \cite{MetricsThesis} - is essentially equal to the nominal $30$ ps jitter on the crystal oscillator. As a result, we identify this jitter on the crystal oscillator as the largest source of error in the WCD after all other design constraints have been accounted for. Therefore, future WCD designs can obtain a greater precision by using a higher quality crystal oscillator with lower jitter.

\section{Calibration}
\label{Calibration}
\Figure[!htb](topskip=0pt, botskip=0pt, midskip=0pt)[width=7in]{\Data}
{Raw data captured by the WCD from measurements of phase-shifted PLL outputs used for calibration. (a): 2.5 ns pulses, 78 ps / phase shift. (b): Zoom of green region in (a). (c): 0.83 ns pulses, 104 ps / phase shift. \label{FData}}

In this section, we begin by examining the features of the DPS-calibration data in Fig. \ref{FData}. We go on to describe the results and technique of the calibration in Fig. \ref{FCalib}. Finally we introduce a novel error-correction algorithm in Fig. \ref{FEC}. 

Figure \ref{FData} (a) and (c) show datasets obtained from dynamic phase-shifted calibration measurements of two different PLL outputs with different frequencies. Horizontal slices correspond to registered states of the chain $\textbf{X}$ after the phase-shift indicated on the vertical direction. Yellow pixels are positions in the CC where the register latched 1, and dark purple 0. Note the unprecedented nature of the data: we capture waveforms at $10^{-12}$ s intervals without error correction or post-processing (`raw data'), from which individual features can readily be examined as in (b).

Shown in Fig. \ref{FData} (a) are 2.5 ns wide pulses generated every 10 ns by a 100 MHz, 25\% duty-cycle PLL output. Every integer multiple of $2\pi$, these pulses align with the rising edge of the 200 MHz capture clock. Hence, as the phase is incremented by $\Delta t=78$ ps down the vertical, the pulses progress horizontally down the chain. The duty cycle is such that a single pulse exists within the carry chain at any given time. This simplifies the calculation of the speed of the rising and falling edges, for which a unique pair of tuples (phase shift, edge carry index) can be obtained from each horizontal slice. From this data, we calculate the speed of the edges and obtain a transfer function $t(x)$, which maps every carry position to a time interval relative to the capture event.

Shown in Fig. \ref{FData} (c) are 0.83 ns pulses occurring every 1.67 ns arising from a 600 MHz, 50\% duty-cycle PLL output. This frequency and duty cycle ensures that multiple pulses exist within the CC at any given time, which simplifies a calculation of the rate at which pulse widths shrink as a function of the position of their falling edge. This phenomenon, known as pulse shrinking, arises due to difference between the speed of the rising and falling edges, and can be seen by observing the pulse widths decrease within a single horizontal slice of (c). For this analysis, we use $\Delta t=104$ ps due to limitations generating incommensurate PLL outputs at high frequencies. 

As can be seen in Fig. \ref{FData} (b), the signal expands as it enters the chain. This is analogous to the change in propagation speed of a wave at the boundary between two media: the first element in the CC is a copy gate, which takes much longer to operate than the subsequent carry elements. This means that the largest differential nonlinearity, or deviation from a constant speed, occurs at the start of the chain. Additionally, bubble errors, which are due to jitter in the arrival of the clock signal at the registers (\textit{i.e}, not all registers latching simultaneously), are clearly visible. These non-idealities - pulse-shrinking, bubble errors, and differential nonlinearities - induce signal distortions.

However, we find that the signal remains faithfully represented even in the presence of these distortions. By way of explanation, we highlight several checks, some of which can be seen from Fig. \ref{FData}. The first such check is to note that the nominal 2.5 ns width of the pulses in (a) are accurately represented along the vertical, demonstrating the stability of the phase-shift. This is also demonstrated by noting that the nominal 10 ns between pulses are clearly visible. Another check (not shown here) is to feed in a signal of constant logic high and ensure that the data are all 1's. This ensures the stability of the registers at the capture frequency.

\Figure[](topskip=0pt, botskip=0pt, midskip=0pt)[width=7in]{\Calib}
{(a): Carry speed calibration over 8 TDLs on 3 FPGAs. (b): Transfer function for a single FPGA and TDL. (c): Pulse widths and linear contraction rate.\label{FCalib}}

We find that the rising and falling edges exhibit a highly linear relationship between edge position and phase shift so that carry rise and fall times can be calculated from a linear fit to the data. This is shown by the black dotted lines in Fig. \ref{FCalib} (a), in which we extract the (phase shift $n\Delta t$, edge carry index $x_{i}$) tuples from data of the kind in Fig. \ref{FData} (a) for the rising (red) and falling (blue) edges. This data is binned over eight TDL locations and three FPGAs, from which we obtain a distribution of carry indices at each phase shift, having standard deviations $\delta x_{i}\sim10$ carries (shown but not visible). These data with errors are fit to straight lines, the reciprocal slopes of which yield the quoted carry times for MLABs  $\tau_{rise}=$\CarryRiseTime{} and $\tau_{fall}= $\CarryFallTime{}. For LABs (not shown) we find \LABCarryRiseTime, \LABCarryFallTime. 

The data in Fig. \ref{FCalib} (a) can be transposed in order to obtain a mapping from carry index $x$ to time interval $t(x)$, realizing full TDC functionality. In the TDC literature, a single TDL would be referred to as a channel and calibrated individually. This is shown in Fig. \ref{FCalib} (b). The uncertainty in the timing $\delta t$ at carry index $x$ is given by the local slope of the mapping $t(x)$ multiplied by the uncertainty in the edge carry index at that time, \textit{i.e}, $\delta t(x)=t'(x)\delta x$, where $t'$ is the derivative of $t(x)$. The quoted uncertainties $\langle \delta t \rangle\sim 30$ ps are the RMS errors (single-shot precision) for each edge over the full dynamic range ($K\tau$). They are of the same magnitude as the jitter of the crystal oscillator \cite{OscillatorDeviceDatasheet}, suggesting the primary source of error is this jitter.

The differences between rise time and fall time cause pulse shrinking, which we calculate explicitly in Fig. \ref{FCalib} (c) from the data shown in \ref{FData} (c). We define a pulse width as the length of any observed number of contiguous 1's. This definition means that bubble errors dominate the distribution and interrupt `real' pulses. This is therefore a simple approximation, and we filter widths below 15 carries to exclude these bubbles. We bin these widths as a function of the position of their falling edge, and show the distributions as violin plots in red (only every 25th plot is shown for clarity). The resulting data is then fit to the dotted line, the absolute value of the slope being the quoted contraction. Finally, we observe no `ultra-wide' bins\cite{KeyIssues} that have been observed in older FPGA process technologies - with the potential exception of the middle of the chain, which may be due to crossing clock-domains.

The fractional difference in rise and fall times is $1-4.91/4.54\approx 8\%$, consistent with the $11\%$ fractional rate of contraction obtained from this basic analysis (for LABs we find closer to $20\%$). This means that the rate of pulse-shrinking is essentially linear. As a consequence, this type of signal distortion can be inverted with a simple error correction algorithm, such as dilating the pulses according to the observed rate of contraction. We examine such an algorithm in the remainder of this section. 

\Figure[!htb](topskip=0pt, botskip=0pt, midskip=0pt)[width=2.5in]{\EC}
{Raw and corrected data from a single state of the CC in \ref{FData} (c). \label{FEC}}

Figure \ref{FEC} shows the result of applying error correction to a single state of the CC from Fig.~\ref{FData}(c). The algorithm fills in all bubble errors and reverses the effects of pulse contraction in software.  We note that the copy element at the start of the CC acts as a low-pass filter that limits the switching time to $\sim50$ ps; thus we isolate all contiguous regions of zeros of width greater than 15 carries and obtain the indices of these regions.  For example, in the raw data shown in Fig. \ref{FEC}, the last such set of zeros is from carry 1033 to 1249. Next, we reverse pulse shortening by reducing the right endpoint of each empty region by a multiplicative factor - 0.95 in the image above (\textit{e.g.}, $1249\rightarrow \left\lfloor 0.95\times 1249 \right\rfloor)$. Finally, we reverse bubble errors by filling in all points which do \textit{not} lie in these regions with logic high. 

We stress that previous TDC error-correction algorithms rely on thermometer encodings, which are not applicable to arbitrary waveforms. However, our algorithm does not account for the expansion at the start of the chain and the extra contraction in the center. More advanced algorithms correcting for these nonlinearities - such as functional transforms rather than constant multiplicative factors - are the subject of future work. 

\Figure[](topskip=0pt, botskip=0pt, midskip=0pt)[width=7in]{\RO}
{(a)-(c): Continuous capture of 3-node ring-oscillator initialized in the unstable (0,0,0) state. Zoom in of (d): in-phase and (e): out-of-phase oscillations.\label{FRO}}
\section{Case Study: Asynchronous Ring-Oscillator}
\label{Results}

Here we briefly demonstrate the DSO capability of the WCD by continuously measuring an asynchronous (unclocked) system of logic gates known as a ring oscillator (RO). ROs are ubiquitous in hardware design and are a natural case study for the WCD. As we will show, our results match those from the literature and allow for precision study of FPGA logic with applications in, \textit{e.g.}, novel attacks on FPGA-based cryptographic primitives.

ROs consist of inverter gates (or nodes) arranged in a unidirectional circle. Each node inverts the state of the previous node in the ring. When the logic gates are in a steady-state, pulses of logic-high travel synchronously around the ring \cite{GHIL}. However, when the states are unsteady, such as in an unclocked system, inversions occur as quickly as the hardware allows. These asynchronous dynamics exhibit truly analog behavior due to collisions between electronic pulses generating intermediary logic values. In these cases, the dynamics of the ring are highly complex and exhibit multi-timescale transients\cite{LOHMANN}.

We capture these dynamics in Fig. \ref{FRO}. Shown in (a),(b),(c) are the continuously captured states of each individual node in a 3-node RO on a single FPGA. The transient and steady behavior discussed above is shown in (d) and (e). The initial state of the RO is set with a multiplexer to the unstable (0,0,0) state before being released and allowed to evolve in time asynchronously. We capture each node every $T=5$ ns, and evaluate the transfer function to truncate the TDL after $T=5$ ns (corresponding to $K=940$) so that we can concatenate the images into a continuous time-series as in (d) and (e).

The observations of these transients reproduce those reported in Ref.~\cite{LOHMANN}, which used a commercial DSO to study the dynamics of a 3-node RO on a Cyclone IV FPGA that were routed off-chip. The steady behavior of (e) matches the theoretical expectations from \cite{GHIL}. Moreover, we find that the mean pulse width is \LogicHighTime{} per inverter gate (see Appendix for method). This is similar to the Cyclone IV 3-node RO inverter gate timescale of $0.28\pm0.01$ ns obtained by \cite{ROSIN} using a commercial DSO. These results confirm the DSO capability of the WCD, as it is expected that the Cyclone V architecture used here has a faster operation time.

As future research, we will perform similar measurements to directly study the properties of other internal FPGA circuitry such as routing delays, PLL jitter, and the rise and fall time of individual logic elements. Such measurements have important consequences for FPGA-based cryptographic primitives such as those described in Ref.~\cite{HBN}, which rely on obfuscation of these properties. As such, the WCD may open new attacks on these security architectures.

\section{Conclusion}
\label{Conclusion}

In summary, we use carry chains and phase-lock loops for the direct capture of digital waveforms and for calibration using dynamic phase-shifting. In doing so, we construct the first ps-scale-resolution FPGA-based measurement system with hybrid digital storage oscilloscope and time-to-digital converter functionality. As a result, the WCD finds application in any scenario where high-precision, ultra-fast measurements are required and a low-cost, off-the-shelf solution is desirable.

There are several advantages to WCD design and DPC. First, both are straightforward: fundamentally the WCD is a buffer, and DPC is a velocity calculation that requires only a single memory read-out. These concepts are easily accessible and, when implemented with the proposed architecture, yield raw-data that is both interpretable and precise. As a result, the WCD is natural for FPGA applications, which aim for ease-of-use.

In contrast, consider the calibrations and requirements of other TDC architectures: bin-realignment, statistical tests, coding-decoding, sorting algorithms, individual carry calibrations, \textit{etc.} \cite{Review2019,CycloneV2ps,Xilinx4ps,Xilinx5ps,XilinxMissingCodeFree,Multiphase,TimingAlignment,WaveUnion,DirectToHist,chenTimeResolutionImprovement2018,wonTimetoDigitalConverterUsing2016}. In Ref.~\cite{CycloneV2ps}, for example (a Cyclone V FPGA TDC), the raw data are randomly distributed and indiscernible - requiring an external hard-processor to implement a sorting algorithm on the encoded distribution of hundreds of randomly aligned clocks.

Second, the WCD design is flexible. Here we report only the novel, minimum number of components. However, an integrated system can easily use the WCD and leverage relevant architectures from DSOs and TDCs for the desired performance or error correction, such as bubble error correction \cite{BubbleCorrection} or a hardware extension of the run-length encoding error correction algorithm developed here.

The flexibility of the design immediately presents several extensions of the WCD. For example, the sampling rate and resource count can be improved by using two registers per carry. DPS can be used to delay the measurement clock and incrementally capture repeated signals in $\Delta t$ intervals. Data transfer off the board can be streamlined using the hard-processor rather than the JTAG interface.

Finally, we stress that analog waveform capture is also possible. Each TDL acts as a 1-bit ADC or comparator as described by $\theta$ in Eq. \ref{WC}. Similar to a flash-ADC, the domain of the input waveform $[V_{min},V_{max}]$ can be partitioned into a series of subsets ranging over the amplitude of the signal $[V_{min},V_{1}),[V_{1},V_{2}),...,[V_{M},V_{max}]$ using external comparators having thresholds corresponding to each sub-interval. The output of each comparator can then be passed into a corresponding TDL. Each of the TDLs would thus trace the associated portion of the analog signal within that sub-domain. The continuous approximation of the amplitude of the waveform could then be recovered from each of the TDLs, yielding an analog WCD. This is the subject of future work.


\appendix{}
\section{}

\subsection{Low-Level Primitives on Cyclone V}
\label{APrim}
The Verilog code used to instantiate our design is listed below and the result of its compilation using the Quartus CAD software is shown in Fig. \ref{FLE}. In a Cyclone V FPGA, the TDLs are instantiated using the low-level Altera Cyclone V primitives \cite{CycloneIICarryPrimitive}. Two of these primitives fit in each cell of a logic-array block (LAB) or memory logic-array block (MLAB). Each cell contains several look-up-tables (LUTs: F0, F1, F2, F3), multiplexers (MUXs), and registers (REGs). A hash specifies the the logic function implemented by manipulating the LUTs, which cause the MPXs to route the input carry CIN through an adder. The adder simply passes the signal to the next MLABCELL and a register. The routing on the register ports are then controlled explicitly and the placement of all logic in each MLAB is set using location constraints. This is accomplished via Tcl commands such as "set\_location\_assignment MLABCELL\_X25\_Y70\_N0 -to name." LABs follow a similar procedure. 

For our ring oscillator experiment, we place the RO compactly within a single MLAB, and use nearby MLABs to create TDLs for each of the three nodes. To determine the pulse width per inverter gate, we instantiate an $N=19$ node RO initialized to the stable state in which the first node is 1 and the rest are 0, and tap only the first node with a CC. We bin the distribution of pulse widths after filtering pulses of width below 15 carries to remove bubbles as in our analysis of pulse shrinking. We calculate the average width and standard deviation, and multiply by the carry time and propagate the error. Finally, we divide the result by $N=19$, as it is well-known that the width of a pulse in an $N$-node ring oscillator is $N$ times the node timescale.

\Figure[h](topskip=0pt, botskip=0pt, midskip=0pt)[width=7in]{\LE}
{Altera Cyclone V low-level primitive implementation of Fig. \ref{FTDL}, showing a typical carry element in a MLABCELL and associated register. Not shown: a special first MLABCELL, occurring at the top of the MLAB block, routes the input signal through a copy element, where it is carried down the chain. Subsequent LEs use a LUT mask of all 0s to eliminate redundant logic and explicitly carry the signal.\label{FLE}}

\begin{lstlisting}[language=Verilog]
/*
This file contains Cyclone V primitives for instantiating a carry chain,
including combinational logic cells, registers,
and a module for the carry chain itself.

The definitions can be found in the quartus\eda\sim_lib\cyclonev_atoms folder
and the functions
cyclonev_lcell_comb
cyclonev_ff

For other FPGAs, follow a similar procedure.
*/

module CarryChain #(parameter N=256) (
    input wire [N-1:0] a,
    input wire [N-1:0] b,
    input wire carryin,
    input wire clk,
    input wire enable,
    input wire clear,
    output wire [N-1:0] regout,
    output wire carryout
  );

//internal wire for passing carries
//one bit larger for initial combinatorial cell
//which passes data_in along carry chain
wire [N:0] c_internal;
//assign first input manually
primitive_carry_in mycarry_in (
  .datac(carryin),
  .cout(c_internal[0]),
);
//internal wire for sum outs
wire [N-1:0] s;
//and rest with loop
genvar i;
generate
for (i=1; i<N+1; i=i+1) begin : gen

  primitive_carry mycarry (
  	.a(a[i-1]),
  	.b(b[i-1]),
  	.cin(c_internal[i-1]), //first carry feeds in, rest daisy chained
  	.cout(c_internal[i]),
  	.s(s[i-1])
  );

  primitive_ff myff(
  	.s(s[i-1]),
  	.clk(clk),
  	.q(regout[i-1]),
    .clr(clear),
    .ena(enable)
  );

end
endgenerate
assign carryout=c_internal[N];
endmodule

//This is the first input copy gate, as you cannot feed to carry port directly
module primitive_carry_in(
  	input wire datac,
    output wire cout
  );

localparam dt = "off";

cyclonev_lcell_comb #(
  .dont_touch(dt)
  ) mycarryin (
                             .dataa(),
                             .datab(),
                             .datac(datac),
                             .datad(),
                             .datae(),
                             .dataf(),
                             .datag(),
                             .cin(0),
                             .sharein(),
                             .combout(),
                             .sumout(),
                             .cout(cout)
);

//This LUT mask passes the datac port to cout
defparam mycarryin.lut_mask=64'h000000000F0F0F0F;

/* the mask works by concatenating the LUTs of the MPXs f0,f1,f2,f3, e.g.
localparam [63:0] mask;
mask[15:0]=f0;
mask[31:16]=f1;
mask[47:32]=f2;
mask[63:48]=f3;
*/
endmodule


//This is the basic carry building block; the selection of ports here matches Fig. 1
module primitive_carry(
  	input wire a,
  	input wire b,
  	input wire cin,
  	output wire cout,
    output wire s
  );

localparam dt = "off";

cyclonev_lcell_comb #(
  .dont_touch(dt)
  ) mycarry (
                             .dataa(),
                             .datab(b),
                             .datac(a),
                             .datad(),
                             .datae(),
                             .dataf(),
                             .datag(),
                             .cin(cin),
                             .sharein(),
                             .combout(),
                             .sumout(s),
                             .cout(cout)
);

//This LUT mask propagates the carry only
defparam mycarry.lut_mask=64'h0000000000000000;

endmodule

//This is the basic register primitive in Fig. 1
module primitive_ff(
  	input wire s,
  	input wire clk,
    input wire clr,
    input wire ena,
    output wire q
  );

cyclonev_ff myff (
    .d(s),
    .clk(clk),
    .clrn(),
    .aload(),
    .sclr(clr),
    .sload(),
    .asdata(),
    .ena(ena),
    .devclrn(),
    .devpor(),
    .q(q)
);

endmodule
\end{lstlisting}

\bibliographystyle{IEEEtran}
\bibliography{IEEEabrv,main}

\begin{IEEEbiography}[{\includegraphics[width=1in,height=1.25in,clip,keepaspectratio]{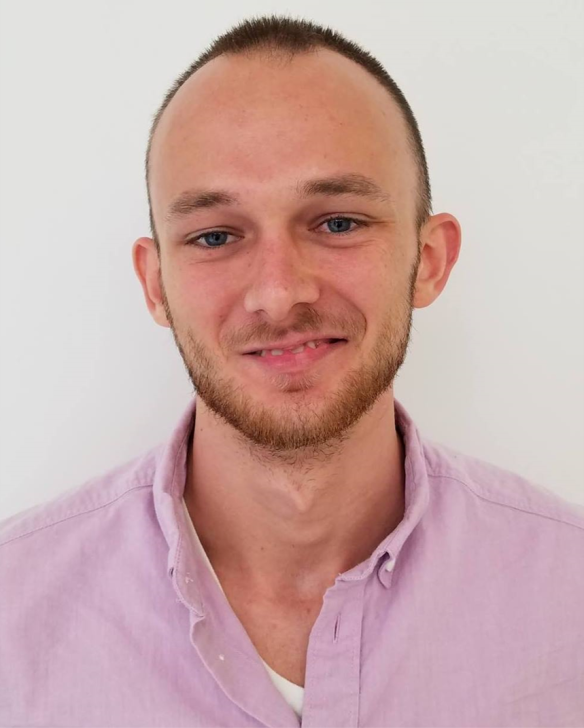}}]{Noeloikeau F. Charlot} 
received the B.S. degrees in physics and biological engineering from the University of Hawai’i at Manoa, Honolulu, HI, USA in 2017, and an M.S. degree in physics from the Ohio State University, Columbus, OH, USA in 2020, where he is currently pursuing a Ph.D. degree in physics. 

Noelo is a research assistant in the QuantInfo lab and CYAN cybersecurity collaboration at OSU. Previously, he interned in the Gravitational Astrophysics lab at NASA Goddard and the Dark Matter Detection lab at UH Manoa. His industrial experience includes bioreactor design and photonics at ProtaCulture, LLC. His research interests include network science, artificial intelligence, ultrafast electronics, and quantum gravity. Noelo is a prior McNair Scholar.
\end{IEEEbiography}

\begin{IEEEbiography}[{\includegraphics[width=1in,height=1.25in,clip,keepaspectratio]{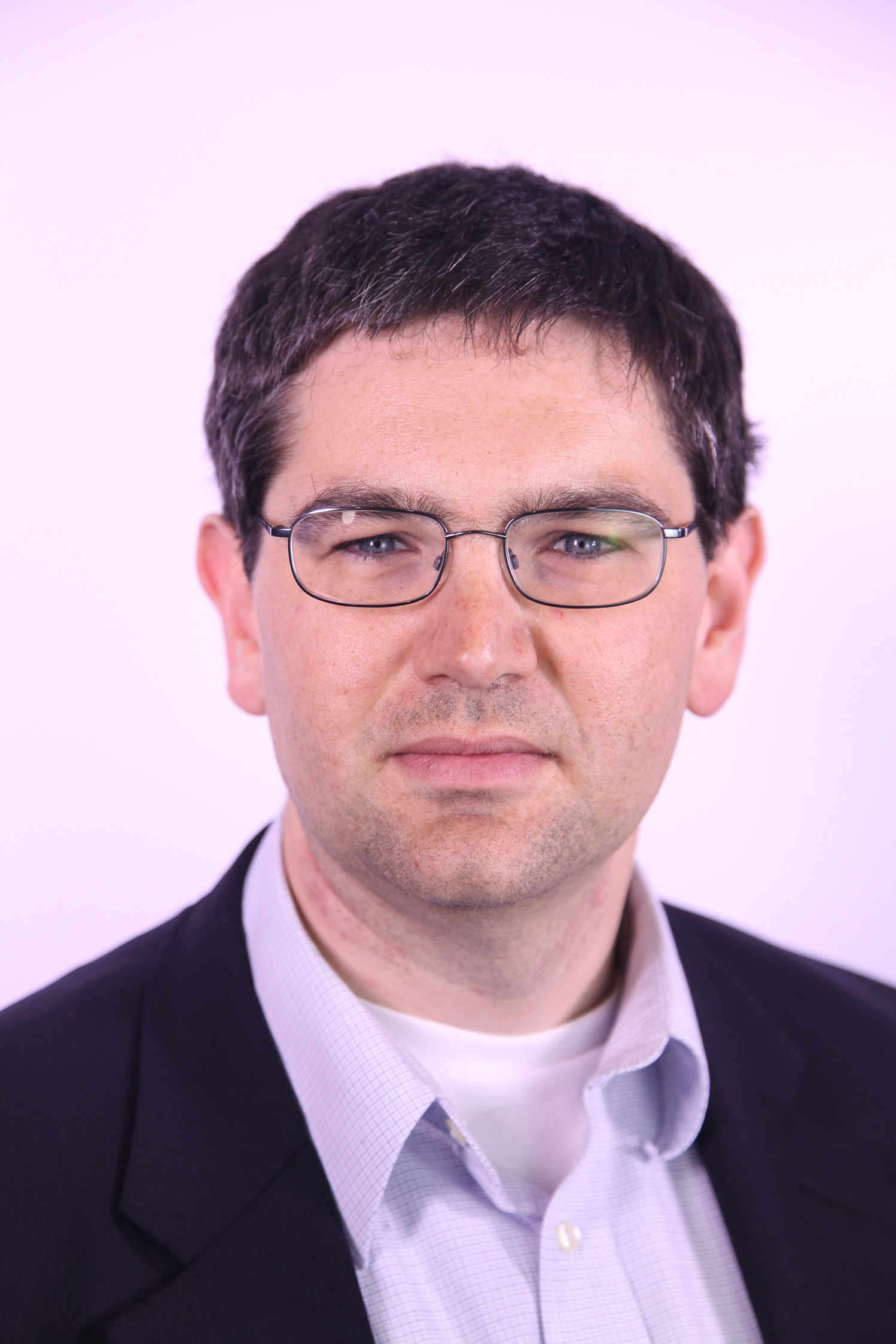}}]{Andrew Pomerance}
Andrew Pomerance  was born in Washington, DC, USA, in 1980. He received the B.S. and M.S. degrees in Electrical and Computer Engineering from Carnegie Mellon University, Pittsburgh, PA, in 2002, and the Ph.D. in Physics from the University of Maryland, College Park, MD, in 2009. 

From 2009 to 2013, he was with Raytheon Applied Signal Technology, Tyson's Corner, VA, USA. He is currently the president of Potomac Research, LLC, Alexandria, VA, USA.  His research is concerned with nonlinear dynamics with applications to machine learning and cryptography.
\end{IEEEbiography}

\begin{IEEEbiography}[{\includegraphics[width=1in,height=1.25in,clip,keepaspectratio]{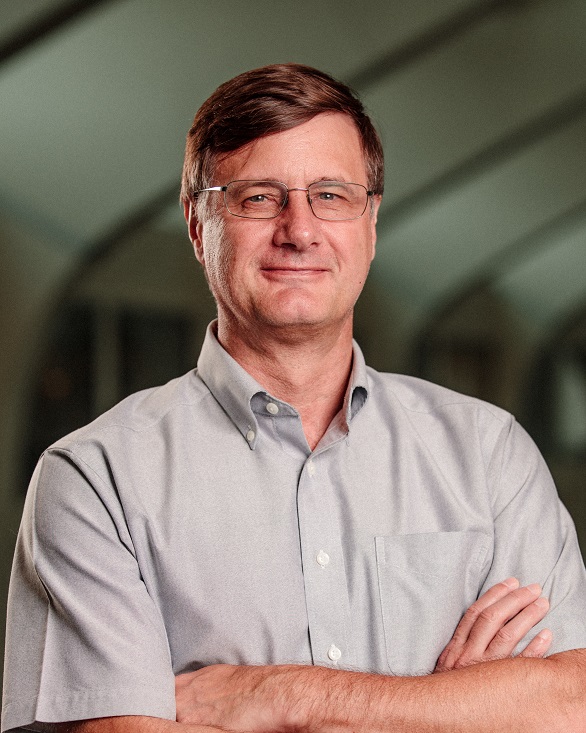}}]{Daniel J. Gauthier}
Daniel J. Gauthier is a Professor of Physics and Electrical and Computer Engineering at The Ohio State University. He received the B.S., M.S., and Ph.D. degrees from the University of Rochester, Rochester, NY, in 1982, 1983, and 1989, respectively. His Ph.D. research on “Instabilities and chaos of laser beams propagating through nonlinear optical media” was supervised by Prof. R.W. Boyd and supported in part through a University Research Initiative Fellowship. From 1989 to 1991, he developed the first CW two-photon optical laser as a Post-Doctoral Research Associate under the mentorship of Prof. T.W. Mossberg at the University of Oregon. In 1991, he joined the faculty of Duke University, Durham, NC, as an Assistant Professor of Physics and was named a Young Investigator of the U.S. Army Research Office in 1992 and the National Science Foundation in 1993.  He was the Robert C. Richardson Professor of Physics at Duke from 2011- 2015, chair of the Duke Physics Department from 2005 – 2011, interim chair in spring 2015, and was a founding member of the Duke Fitzpatrick Institute for Photonics. He moved to The Ohio State University in 2016. His research interests include: reservoir computing, synchronization and control of the dynamics of complex networks in electronic and optical systems, quantum communication, and nonlinear quantum optics. Prof. Gauthier is a Fellow of the Optical Society of America and the American Physical Society.
\end{IEEEbiography}

\EOD

\end{document}